\documentclass[preprint,aps,showpacs,nofootinbib]{revtex4}

\usepackage{epsfig,amssymb,amsmath}

\newcommand{\beq}{\begin{equation}}
\newcommand{\eeq}{\end{equation}}
\newcommand{\bea}{\begin{eqnarray}}
\newcommand{\eea}{\end{eqnarray}}
\newcommand{\bear}{\begin{array}}
\newcommand {\eear}{\end{array}}
\newcommand{\bef}{\begin{figure}}
\newcommand {\eef}{\end{figure}}
\newcommand{\bec}{\begin{center}}
\newcommand {\eec}{\end{center}}

\def\REF#1{(\ref{#1})}
\def\GEV#1{10^{#1}{\rm\,GeV}}

\def\lrf#1#2{ \left(\frac{#1}{#2}\right)}
\def\lrfp#1#2#3{ \left(\frac{#1}{#2} \right)^{#3}}

\newcommand{\gp}{\gamma^\prime}

\newcommand{\dnf}{\Delta N_{\rm eff}}

\begin{document}
\draft
\tighten
\preprint{RIKEN-MP-67}
\preprint{TU-930}
\title{\large \bf
A Parallel World in the Dark
}
\author{
    Tetsutaro Higaki$^{\,a,\star}$\footnote[0]{$^\star$ email: tetsutaro.higaki@riken.jp},
    Kwang Sik Jeong$^{\,b,\ast}$\footnote[0]{$^\ast$ email: ksjeong@tuhep.phys.tohoku.ac.jp},
    Fuminobu Takahashi$^{\,b,\dagger} $\footnote[0]{$^\dagger$ email: fumi@tuhep.phys.tohoku.ac.jp}
    }
\affiliation{
    {\it $^a$ Mathematical Physics Lab., RIKEN Nishina Center, Saitama 351-0198, Japan}\\
    {\it $^b$ Department of Physics, Tohoku University, Sendai 980-8578, Japan}
    }

\vspace{2cm}

\begin{abstract}
The baryon-dark matter coincidence  is a long-standing issue.
Interestingly, the recent observations suggest the presence of dark radiation,
which, if confirmed, would pose another coincidence problem
of why the density of dark radiation is comparable to that of photons.
These striking coincidences may be traced back to
the dark sector with particle contents and interactions that are quite similar,
if not identical, to the standard model: a dark parallel world.
It naturally solves the coincidence problems of dark matter and dark radiation,
and predicts a sterile neutrino(s) with mass of ${\cal O}(0.1 - 1)$\,eV,
as well as self-interacting dark matter made of the counterpart of ordinary baryons.
We find a robust prediction for the relation between the abundance of dark
radiation and the sterile neutrino, which can serve as the smoking-gun evidence
of the dark parallel world.
\end{abstract}

\pacs{}
\maketitle


\section{Introduction}
The observed dark matter density is about five times
larger than the baryon density,
\beq
\label{coin1}
\Omega_{\rm DM} \; \simeq \; 5\, \Omega_B.
\eeq
If  the baryon and the dark matter have a totally different origin,
it is a puzzle why they happen to have the density in a similar size.
This is known as the baryon-dark matter coincidence problem, and
many solutions have been proposed
(See e.g. Refs.~\cite{Fujii:2002aj,Seto:2007ym,Kaplan:2009ag,Shoemaker:2009kg,
Cheung:2011if,Doddato:2011hx,Kasuya:2012mh,Jeong:2013axf}.)
Intriguingly,  the recent SPT and WMAP observations suggest the presence of dark radiation with
$N_{\rm eff} = 3.71 \pm 0.35$~\cite{Hou:2012xq} and $3.84 \pm 0.40$~\cite{Hinshaw:2012fq},
respectively.
If confirmed, it would pose another coincidence problem of
why the density of dark radiation is comparable to that of photons;
\beq
\label{coin2}
\Omega_{\rm DR} \;\simeq\; 0.23\, \dnf\, \Omega_{\gamma},
\eeq
where $\dnf \equiv N_{\rm eff} - 3.046$, and $\Omega_{\rm DR}$ and $\Omega_\gamma$ denote
the density parameter of dark radiation and photons, respectively.\footnote{
See Refs.~\cite{Chun:2000jr,Ichikawa:2007jv,Nakayama:2010vs,Fischler:2010xz,Hasenkamp:2011em, Menestrina:2011mz,
Kobayashi:2011hp,Hooper:2011aj,Jeong:2012np, Cicoli:2012aq,Choi:2012zna,Graf:2012hb,Hasenkamp:2012ii,
Bae:2013qr,Graf} for possible explanations of dark radiation.
}
These coincidences \REF{coin1} and \REF{coin2} would be striking enough to make one
to reconsider what dark matter and dark radiation are.

The two coincidences may be traced back to the dark sector with particle contents and
interactions that are quite similar, if not identical,
to the standard model (SM). That is, the dark sector is a parallel world of the SM,
with slightly different couplings and (light) fermion masses.
The dark parallel world naturally solves the coincidence problems of dark matter as well as
dark radiation.
It also predicts self-interacting dark matter~\cite{Spergel:1999mh} made of para-baryons\footnote{
Throughout this paper we denote the counterpart of the SM particles by adding a prime or a
prefix para- or simply p-.
For instance the counterpart of a photon, $\gamma^\prime$, is called para-photon or simply p-photon.
},
as well as dark radiation and sterile neutrinos which are nothing but the counterparts of
photons and neutrinos. 

A replica of the SM was considered in many literatures.
In a seminal paper \cite{Lee:1956qn}, Lee and Yang proposed that there must exist a mirror world,
which compensates the mirror (or parity) asymmetry of the weak interactions of the SM. The mirror world
has been studied from both phenomenological and astrophysical aspects (see
e.g. Refs.~\cite{Blinnikov:1982eh,Kolb:1985bf,Khlopov:1989fj,Foot:1991bp, Hodges:1993yb,Berezhiani:2005ek, Okun:2006eb}
and references therein), and many papers studied the mirror
world with an exact  $Z_2$ symmetry that interchanges the mirror world and the SM sector.
In the presence of such  $Z_2$  symmetry,  we can directly apply our knowledge of the SM to the mirror world,
enabling us to make rather robust predictions about the property and evolution of the mirror sector.
In particular, the mirror baryons are a plausible candidate for dark matter.
However, in order to satisfy the big bang nucleosynthesis constraint on additional light degrees of freedom,
one needs to introduce e.g. asymmetric reheating~\cite{Berezhiani:1995am}.
There are many other issues such as structure formation.  In another class of models called
``asymmetric mirror world",
although the couplings in the mirror sector are  identical to those in the SM sector,
the breaking scale of the mirror electroweak symmetry is assumed to differ from the SM value by
a factor $\zeta$, which eases several difficulties of the symmetric mirror
world~\cite{Mohapatra:1996yy,Mohapatra:2001sx}.
Also, additional couplings between the two sectors other than gravitation were often introduced,
such as the neutrino mixings~\cite{Foot:1993yp,Berezhiani:1995yi,Berezhiani:1995am}, a kinetic mixing with the hidden
photon~\cite{Holdom:1985ag,Glashow:1985ud,Carlson:1987si,Foot:1991bp},
and a Higgs portal~\cite{Barbieri:2005ri}, etc.

In our paper, we only require that the parallel world has the same gauge structure and
particle content as the SM sector ones, but the parameters in the Lagrangian can
be slightly different.\footnote{We shall comment on a case in which the particle content of the parallel
sector is different from that of the SM.}
Especially the mass of light fermions (i.e., up and down quarks, electrons and neutrinos)  could be different
by a factor of several, or maybe one order of magnitude with respect to the SM values.
We  also focus on a case in which the parallel world and the SM sector
are coupled only through gravitation.  We shall derive a non-trivial relation
of the abundance of dark radiation $\dnf$
and the effective number of massive sterile neutrinos $N_s$,
which will serve as the smoking-gun evidence of the dark parallel world.
Note that, although the amount of dark radiation itself cannot be definitely predicted in our
framework\footnote{
The amount of additional effective number of neutrinos
was estimated in terms of $\zeta$ in Ref.~\cite{Mohapatra:2000qx}, assuming massless mirror neutrinos.
},
the ratio of the abundances of dark radiation to sterile neutrino is simply determined by the number
of light degrees of freedom.

The rest of this paper is organized as follows. In Sec.~\ref{sec:2} we derive the relation
between $\dnf$ and $N_s$ and also study the baryon-dark matter coincidence problem.
We discuss the implications from string theory in Sec.~\ref{sec:3}, and the last section
is devoted to discussion and conclusions.

\section{A dark parallel world}
\label{sec:2}
In this section we discuss cosmological and astrophysical aspects of the dark parallel world.
As we mentioned in the previous section, we assume that the parallel world  is quite similar
to the SM sector, since otherwise it would be difficult to explain the similarity in the abundance
of baryon and dark matter and that of the ordinary and dark radiation.
The two sectors are assumed to interact with each other only through gravitation.
Other interactions such as a kinetic mixing between U(1)'s, a Higgs portal,
and the active-sterile neutrino mixings etc are possible and would offer an interesting probe
of the parallel world, but we do not consider them here since they are
quite model-dependent.

In many literature, it is assumed that there is an exact $Z_2$ symmetry
which interchanges the SM sector with the parallel world. However,
we do not impose such an exact $Z_2$ symmetry. This is because of the apparent disparity
between dark and ordinary matter;
for instance, the radiative cooling of gas is a crucial ingredient of galaxy formation,
while the dark matter should not cool off by radiating p-photons. Such difference depends
on the detailed properties of p-nucleons as well as the p-electron,
and it could arise as 
a result of a slightly 
different choice of the parameters in the Lagrangian.
We shall come back to this issue later.
Due to our lack of knowledge of the parallel world, it is difficult to
study detailed properties of those particles unless some exact symmetry is imposed.
Nevertheless, the apparent similarity in the cosmic
abundances strongly suggests that the disparity of the parameters between the two sectors is not
significant.

Our key question is if we can find any robust predictions of this framework.
We observe that (i) the existence of a sterile neutrino(s) and (ii)
self-interacting dark matter made of para-baryons;  these two points are robust predictions that
are mostly independent of details of the parallel sector.  We will show that there is
a robust prediction for the relation between the abundance of dark radiation and the sterile neutrino,
which can serve as a smoking-gun evidence of the parallel world.

\subsection{Dark radiation and sterile neutrino}
At very early times, the Universe was dominated by the ordinary radiation.
Because of the apparent similarity, we expect that the parallel sector was also thermalized.
We here simply assume that the inflaton decay reheated both the SM and parallel sectors.
The similarity in the abundance of baryon and dark matter suggests that the
two temperatures are comparable to each other, but the two sectors are
thermally decoupled since we assume only gravitational interactions between them.

The counterparts of photons and massless neutrinos contribute to dark radiation, which is
characterized by $\Delta N_{\rm eff}$ as
\beq
\dnf \;\equiv \; \frac{\rho_R^\prime}{\rho_{\nu}^{({\rm th})}}
\eeq
with
\beq
\rho_{\nu}^{({\rm th})} \;=\; \frac{7 \pi^2}{120}\left(\frac{4}{11}\right)^{4/3} T_\gamma^4
\eeq
where $\rho_R^\prime$ and $\rho_{\nu}^{({\rm th})}$ denote the energy density of the extra
radiation and that of one massless neutrino species with a thermal distribution, respectively,
and $T_\gamma$ is the photon temperature after the electron-positron annihilation.
The $\dnf$ is understood to be evaluated during the CMB epoch, unless otherwise stated.

If some of the p-neutrinos are massive, their effect on the CMB and large-scale structure
can be parameterized by their effective number $N_s$ and the mass $m_s$.
The contribution to $N_s$ from one massive p-neutrino species is given by
\beq
N_s \;\equiv\; \frac{n_\nu^\prime}{n_\nu^{({\rm th})}}
\eeq
with
\beq
n_{\nu}^{({\rm th})} \;=\; \frac{6 \zeta(3)}{11 \pi^2} T_\gamma^3,
\eeq
where $n_\nu^\prime$ and $n_{\nu}^{({\rm th})}$ denote the number density of the massive p-neutrino
and that of one species of ordinary neutrinos, respectively.
The p-neutrino mass $m_s$ is unknown, but we assume that it is of
order ${\cal O}(0.1 - 1)$\,eV, because of the expected similarity between the visible and parallel sectors.
In the case of normal hierarchy (NH) for the neutrino masses,
we can approximate that only the heaviest p-neutrino is massive, while the others are massless.
In the case of inverted hierarchy (IH), on the other hand, the two heaviest p-neutrinos are treated
as massive.

In the present Universe, we expect that the light degrees of freedom in the parallel sector are
p-photons and p-neutrinos.
Their contributions to $\Delta N_{\rm eff}$ and $N_s$ depend on the ratio of
their temperatures after the p-electron and p-positron annihilation phase,
\bea
r \equiv \frac{T^\prime_\nu}{T^\prime_\gamma},
\eea
where $r=(4/11)^{1/3}$ if the annihilation takes place after the p-neutrino decoupling as in
the SM sector, and $r=1$ otherwise.
The p-neutrons and p-photons will have the same temperature if the p-electron is several
times heavier than the electron.
Such an assumption may not be entirely unreasonable because
some of the SM parameters, especially the light fermion masses,
may be finely tuned for the existence of life.
 For instance, the closeness of the electron mass and
the  neutron-proton  mass difference may be as a result of such fine-tuning.
It is also known that the electron mass tends to be lower than the typical
prediction of some flavor models.
More importantly, a heavier p-electron mass will reduce the bremsstrahlung,
which may help to account for the different cooling property of dark and ordinary matter.

During big bang nucleosynthesis (BBN), the extra radiation increases the expansion rate,
which subsequently enhances the $^4$He abundance.
The measurements of the primordial $^4$He mass have a somewhat checkered history since
it is very difficult to estimate systematic errors.
The analysis by Izotov and Thuan  gives $\dnf^{\rm BBN} = 0.68 ^{+0.8}_{-0.7}$ at
$2 \sigma$~\cite{Izotov:2010ca}, and we take this as an upper bound on the additional
light degrees of freedom:
\beq
\dnf^{\rm BBN} \;\lesssim\; 1.48 ~~(2 \sigma),
\label{bbn}
\eeq
where the superscript ``BBN" means that $\dnf$ is evaluated during the BBN epoch.
Since p-photons as well as three p-neutrinos contribute to the extra radiation during BBN,
we obtain
\beq
\dnf^{\rm BBN} \;=\; \frac{8+21r^4}{7} \lrfp{11}{4}{\frac{4}{3}} \lrfp{T_\gp}{T_\gamma}{4}.
\label{Neff_bbn}
\eeq
Using \REF{bbn} and \REF{Neff_bbn}, we have
\beq
\label{bbnT}
\frac{T_\gp}{T_\gamma} \,\lesssim\,
\left\{
\bear{cl}
0.55 & {\rm ~~for~~} r=1 \\
0.67 & {\rm ~~for~~} r=(4/11)^{1/3}
\eear
\right..
\eeq
Thus, the temperature of the parallel world must be lower than about half
of the visible sector temperature. Such difference can be explained if
the inflaton coupling to the SM sector is a few times larger than the coupling to the parallel world.

Now we estimate $\dnf$ and $N_s$ at the CMB epoch:
\bea
\label{CMB-Neff}
\dnf^{} &=&
\frac{8+7(3-n)r^4}{7} \lrfp{11}{4}{\frac{4}{3}}\lrfp{T_\gp}{T_\gamma}{4}
\,\lesssim\,
\frac{1.48 (8+7(3-n)r^4)}{8+21 r^4},
\\
\label{CMB-Ns}
N_s &=& \frac{11}{4} n r^3 \lrfp{T_\gp}{T_\gamma}{3}
\,\lesssim\,
\frac{5.77 nr^3}{(8+21r^4)^{3/4}},
\eea
for $n$ massive and $(3-n)$ effectively massless p-neutrino species,
where the inequalities come from the BBN bound (\ref{bbn}).
In the case that the p-neutrinos and p-photons have the same temperature,
we obtain
\bea
\bear{llll}
\dnf^{} \lesssim &
\left\{
\bear{l}
1.12 \\
0.77
\eear
\right.,\,\,
&
N_s \lesssim &
\left\{
\bear{ll}
0.46 & \,\,\quad\mbox{for NH} \\
0.92 & \,\,\quad\mbox{for IH}
\eear
\right..
\eear
\eea
Here the neutrino mass hierarchy means the one in the parallel sector, assuming the same
hierarchy in the SM and parallel sector.
It will be straightforward to apply the relations (\ref{CMB-Neff}) and (\ref{CMB-Ns})
to the case with a different mass hierarchy between the two.
On the other hand, $\Delta N^{}_{\rm eff}$ and $N_s$ are bounded to be
\bea
\bear{llll}
\dnf^{} \lesssim &
\left\{
\bear{l}
1.28 \\
1.08
\eear
\right.,\,\,
&
N_s \lesssim &
\left\{
\bear{ll}
0.3 & \,\,\quad\mbox{for NH} \\
0.6 & \,\,\quad\mbox{for IH}
\eear
\right.,
\eear
\eea
for $r=(4/11)^{1/3}$, which is the case when the p-neutrinos decouple from the thermal
background before the p-electron and p-positron annihilation.

\begin{figure}[t]
\begin{center}
\begin{minipage}{16.4cm}
\centerline{
{\hspace*{0cm}\epsfig{figure=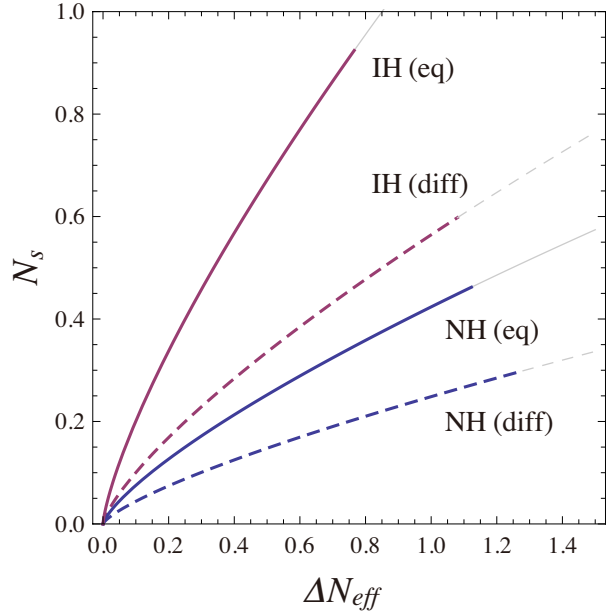,angle=0,width=8cm}}
}
\caption{
The relation of the abundance of dark radiation $\dnf$ and the effective number
of massive sterile neutrinos $N_s$.
The two solid lines correspond to the relation \REF{teq} for the normal (blue, lower) and inverted
(red, upper) hierarchy, respectively. The relation \REF{tdiff} is similarly shown as dashed lines.
The dark parallel sector leads to $\Delta N^{\rm BBN}_{\rm eff}>1.48$ along the gray lines.
}
\label{fig1}
\end{minipage}
\end{center}
\end{figure}

Since we do not know the exact temperature of p-photons, $T_\gp$, we cannot
predict the exact values of $\dnf$ and $N_s$.
Nonetheless, the relation between $\dnf$ and $N_s$ is simply determined by the number
of light degrees of freedom (i.e. p-photon and p-neutrinos)
and the p-neutrino mass hierarchy.
Thus we obtain a rather robust prediction for the relation,
\bea
\label{pred}
N^{4/3}_s = \kappa\, \dnf^{},
\eea
with $\kappa$ given by
\bea
\kappa = \frac{7 n^{4/3}r^4}{8+7(3-n)r^4}.
\eea
This is the main result of this paper.
The value of $\kappa$ is fixed by the p-neutrino mass hierarchy, with a simple thermodynamic
argument.
To be concrete, for the parallel sector with $r=1$, we predict
\beq
\label{teq}
\kappa \simeq \left\{
\bear{cl}
0.32  & \quad\mbox{for NH} \\
1.2   & \quad\mbox{for IH}
\eear
\right..
\eeq
Similarly, the parallel sector leads to
\beq
\label{tdiff}
\kappa \simeq \left\{
\bear{cl}
0.16  & \quad\mbox{for NH} \\
0.47  & \quad\mbox{for IH}
\eear
\right.,
\eeq
if the ratio between p-neutrino and p-photon temperatures is $r=(4/11)^{1/3}$ as in the SM sector.
If such a relation is confirmed by future observations, it will be a smoking-gun evidence of
the dark parallel world.

In Fig.~\ref{fig1}, we show the relations \REF{teq} and \REF{tdiff} of
the effective number of massive sterile neutrinos $N_s$ and the abundance of
dark radiation $\dnf$.
It depends on the sterile neutrino mass scale $m_s$,
whether or not such a relation can be checked by observations.
If $m_s$ is much lighter than $0.1$\,eV, it will be challenging, but if it is of ${\cal O}(0.1-1)$\,eV,
we may be able to confirm the relation.

\subsection{Baryon and dark matter}
The p-baryon is the leading dark matter candidate in our framework. The p-baryon asymmetry is generated
by the p-baryogenesis mechanism similar to the one responsible for the baryon asymmetry in the SM sector.
Thus, these two asymmetries are naturally comparable to each other.
After the p-QCD phase transition, the p-quarks form p-hadrons, and the lightest one will have a mass
of order the dynamical scale of p-QCD, $\Lambda_{\rm QCD}^\prime$. The lightest p-hadron is considered
to be stable in a cosmological time scale because of the p-baryon number conservation, and thus a plausible
candidate for dark matter.
The baryon-dark matter coincidence is a natural outcome of this scenario.

As mentioned at the beginning of this section,
there is clear disparity between baryon and dark matter. The gas made of ordinary matter cools and condenses
in the dark matter's potential, while the dark matter does not cool and it forms  tri-axial haloes.
Due to the lack of our knowledge about the parallel world,
we are not able to predict or explain the different behavior of baryon and dark matter from the first principle,
and so, here we content ourselves with the following simple argument.

Suppose that p-proton is lighter than p-neutron and it is stable in a cosmological time scale.
After the recombination, p-electrons and p-protons  become bound to form electrically neutral
p-hydrogen atoms.  However, the p-hydrogen atoms are considered to have too large scattering cross section
of order $10^{-16} {\rm cm}^2$ at small relative velocities, if all the mass scales in the parallel sector are
same as in the SM. This problem can be circumvented by increasing the breaking scale of the weak interaction
in the parallel sector~\cite{Mohapatra:1996yy,Mohapatra:2001sx}.
Also,  the bremsstrahlung process in ionized gas will be suppressed if the p-electron is heavier.

Another possibility is that p-neutron is the lightest p-hadron. As we shall see below, the scattering cross section
becomes significantly smaller than the previous case. Also, heavier p-nuclei may be unstable if the p-proton mass
is so heavy that the mass difference between p-neutron and p-proton exceeds the typical binding energy of p-nuclei.
Even if this is not the case,
it sensitively depends on the p-proton lifetime how much heavier p-nuclei are created in the nucleosynthesis.
Indeed, in the SM sector, the closeness of the electron mass and the neutron-proton mass difference makes the neutron lifetime
significantly long. If the mass difference were twice larger, the neutron lifetime would be about $50$ times shorter.
Thus, it is possible that the p-proton lifetime is so short that any heavier p-nuclei are not created even if some of them are
stable. Without, or with a tiny amount of charged p-nuclei and p-electrons, the dark matter cannot cool off leading to a tri-axial halo shape.
Since no molecules are formed in this case, there is no star formation in the parallel sector. We emphasize here that
such a drastic change of behavior of dark matter is possible if the light fermion masses in the parallel sector are
slightly different from the SM values.  In the following we assume the latter case in which the p-neutrons constitute dark matter.

There is one robust prediction about dark matter made of p-neutrons,  which does not depend on
the detailed property of the p-hadrons.
That is, the p-neutron dark matter has self-interactions through the counterpart
of the strong interaction. The cross section of the p-neutron dark matter can be inferred from
the scattering cross section of neutron-neutron collisions at low energy,  $\sigma_{nn} \simeq 4\pi a_{nn}^2 \simeq 
45 $\,barn, where $a_{nn} = -18.9 \pm 0.4\,$fm is the neutron-neutron scattering length~\cite{Chen:2008zzj}.
Thus, if the parallel sector is identical to the SM, the ratio of the cross section to the mass is 
given by $\sigma_{\rm DM}/m_{\rm DM} \sim \sigma_{nn}/m_n\, \sim 48\,{\rm barn/GeV} 
\sim 30\,{\rm cm^2/g}$. For a p-QCD scale several times greater than the QCD scale, 
the ratio is considered to be suppressed by a few order of magnitude.\footnote{
The neutron mass is simply determined by the QCD scale, while the neutron-neutron scattering length 
non-trivially depends on the QCD scale as well as masses of pions and other mesons~\cite{Beane:2013br}. 
However, the scattering length is considered to decrease as the QCD scale and light quark masses increase.
}

%

Such self-interacting dark matter was proposed as a possible solution to the discrepancy
between observations and simulations known as central density problem~\cite{Spergel:1999mh}.
The upper bound on $\sigma_{\rm DM}/m_{\rm DM}$ was obtained by several groups as
$\sigma_{\rm DM}/m_{\rm DM} \lesssim 0.1$~\cite{Yoshida:2000uw},
$0.3$~\cite{Gnedin:2000ea}, and
$ 0.02 \,{\rm cm^2/g}$~\cite{MiraldaEscude:2000qt},
comparing the numerical simulations with observations. The constraint from the merging clusters known as
the Bullet Cluster was derived in Ref.~\cite{Randall:2007ph} as
$\sigma_{\rm DM}/m_{\rm DM} \lesssim 0.7 \,{\rm cm^2/g}$.
Recently, the observational constraint on the self-interacting dark matter was carefully re-considered
in Refs.~\cite{Rocha:2012jg,Peter:2012jh},  taking account of various effects which were overestimated
in the previous works and it was found that $\sigma_{\rm DM}/m_{\rm DM}  = 1  \,{\rm cm^2/g}$ is likely inconsistent with
halo shape of the observed clusters, while  $\sigma_{\rm DM}/m_{\rm DM}  = 0.1  \,{\rm cm^2/g}$
is still allowed.
According to Ref.~\cite{Zavala:2012us}, the self-interacting dark matter with 
 $\sigma_{\rm DM}/m_{\rm DM}  \simeq 0.6  \,{\rm cm^2/g}$ can account for the central densities
 of the Milky Way dwarf spheroidals, while the dark matter with $\sigma_{\rm DM}/m_{\rm DM}  = 0.1  \,{\rm cm^2/g}$ 
 can hardly be distinguished from the collisionless cold dark matter. Thus, the self-interacting dark matter with
 the cross  section of  $\sigma_{\rm DM}/m_{\rm DM}  = {\cal O}(0.1)   \,{\rm cm^2/g}$ has  interesting implications for the 
central density problem. 

In order for  the p-neutron cross section to fall in this range, the p-QCD scale (and/or the light p-quark masses) must
be several times larger than the SM values.  This implies that, if the p-baryon number
 is equal or very close to the ordinary baryon number, the dark matter density can be several times greater than the baryon density, 
 which nicely fits the observations \REF{coin1}.  On the other hand, if the p-QCD scale is more than one order of magnitude higher than
 the QCD scale, 
 the self interactions will likely be too small to be distinguished from the collisionless cold dark matter.

Now let us consider a more specific mechanism for baryogenesis.
The similarity in the abundance of baryon and dark matter suggests that
the successful baryogenesis should not rely on fine-tuning of the parameters,
since we do not impose an exact $Z_2$ symmetry between the SM and parallel world
and the same degrees of fine-tuning in the two sectors are unlikely unless
there is a novel physical mechanism for it.
One particularly attractive mechanism that offers an explanation of the observed baryon asymmetry is
leptogenesis~\cite{Fukugita:1986hr}.
Once the Universe was heated to sufficiently high temperature, the leptogenesis
takes place automatically through out-of-equilibrium decays of the heavy right-handed
neutrinos.  We therefore consider thermal leptogenesis in the seesaw mechanism of the
neutrino masses~\cite{seesaw}.

To simplify our discussion, we assume a hierarchical mass spectrum for the right-handed
neutrinos, $M_1 \ll M_2, M_3$, where $M_{1,2,3}$ denotes the right-handed neutrino mass.
The baryon asymmetry in the SM sector generated by the decay of the lightest right-handed neutrino
 $N_1$ via thermal leptogenesis is roughly given by
\beq
\frac{n_B}{s}\;\sim\; 10^{-10} \lrf{k}{0.1} \lrf{M_1}{\GEV{10}}
\lrf{m_{\nu 3}}{0.05 {\rm\,eV}} \delta_{CP},
\eeq
where $k$ represents a wash-out effect as well as the production efficiency of $N_1$,
$M_1$ denotes the mass of the lightest right-handed neutrino, and
$m_{\nu 3}$ is the mass of the heaviest light neutrino, and $\delta_{CP}$ represents the
CP violating phase. Here and in what follows we assume normal mass hierarchy
for the neutrino and p-neutrino.  Similarly, the p-baryon asymmetry is\footnote{
Precisely speaking, the efficiency factor $k'$ is generically different from $k$. 
The thermal leptogenesis in the parallel sector is obtained by multiplying a decay parameter
$K$ with a factor of $(T_\gamma^\prime /T_\gamma)^2 \simeq 0.3 - 0.45$ in the usual thermal 
leptogenesis (see Ref.~\cite{Buchmuller:2004nz}). This is because, for the parallel world, the visible
radiation increases the Hubble parameter and $K \propto 1/H$. 
This however does not change the value of $k'$ significantly for the parameters considered here. 
}
\beq
\frac{n_B^\prime}{s}\;\sim\; 10^{-10}
\left(\frac{T^\prime_\gamma/T_\gamma}{0.5}\right)^3
\lrf{k^\prime}{0.1}
\lrf{M_1^\prime}{\GEV{10}} \lrf{m_{s}}{0.4 {\rm\,eV}} \delta_{CP}^\prime.
\eeq
Note that the p-baryon abundance is normalized to the visible entropy density,
and so it is suppressed by the ratio of the temperature of the two sectors.
Thus, the p-baryon (i.e. dark matter) density
would be about one order of magnitude smaller than the baryon abundance if all the parameters
of the parallel sector are same as the SM ones. Even for $\Lambda_{\rm QCD}^\prime =  (3 \sim 5) \times \Lambda_{\rm QCD}$,
we need to enhance the dark matter abundance by one order of magnitude in order to satisfy \REF{coin1}.

There are several options (or some combination of them) for enhancing the dark matter density\footnote{
See also \cite{Cui:2011wk}.
}:
\begin{enumerate}
\item $ |\delta_{CP}| < |\delta_{CP}^\prime|$
\item $ M_1 < M_1^\prime$
\item $ m_{\nu 3} < m_s$
\item $\Lambda_{\rm QCD} < \Lambda_{\rm QCD}^\prime$
\end{enumerate}
Among the above possibilities,  the third one is particularly interesting, because
it may be possible to detect the effect of the sterile neutrino imprinted in the CMB temperature
anisotropy and the large scale structure data, which enables us to check our prediction (\ref{pred}).

\section{Implications from the string theory}
\label{sec:3}

In this section, we shall consider possibilities on the embedding of the SM and its parallel
world into the string theory.
Since the theory can be a quantum unified theory including gravity,
it is worthwhile to discuss it.
As well-known, string theories require six extra spatial dimensions in addition to
the observed four spacetime dimensions.
The extra dimensions can possess rich structures although
the spaces should be compactified to such a small size that we do not observe them directly.
Furthermore, by involving D-branes in type II string theories it is possible to introduce
gauge theories on them, which can lead to the SM and its parallel world at low energy scales.
Thus it may be natural to have parallel world.

Consider a compact Calabi-Yau (CY) space involving D-branes in IIB/F theory as an example
\cite{Blumenhagen:2006ci, Denef:2008wq, Weigand:2010wm, Maharana:2012tu}.
Then one may find two relevant things, which
are the presence of lots of cycles and the tadpole conditions on the CY space.

First, the CY space can contain many cycles or singularities.
For instance, one may find $h^{1,1}({\rm CY}) > 1$ and $h^{2,1}({\rm CY}) \sim{\cal O}(100)$,
and there would be a cycle which is homologous to another one.
Therefore, when two stacks of D-branes are put on such two different n-cycles of
$\Sigma_n$ and $\Sigma'_n$ separately,
one stack may look like a hidden dark sector from the other:
\beq
\int_{\Sigma_n} d^n y \int d^4 x {\cal L}_{\rm SM}
+ \int_{\Sigma'_n} d^n y \int d^4 x {\cal L}_{\rm para}.
\eeq
Two sectors interact with each  other at least via gravitation.
Let us focus on the model-building of the SM and dark sector
under such a situation.\footnote{
We will not tell about K-theory \cite{Witten:1998cd}
and Freed-Witten anomalies \cite{Freed:1999vc}
on D-brane configurations because it requires more concrete discussion
of model-buildings.
}
For instance, world volume fluxes on two stacks of D7-branes wrapping on two 4-cycles
can lead to copies of chiral fermions (generation)
via index of Dirac operators in the respective sectors \cite{Blumenhagen:2000wh}.
The flavor structure can originate from integrands of three zero-mode wavefunctions
which are localized on the respective branes
\cite{Cremades:2004wa},
and depend also on non-perturbative effects \cite{Blumenhagen:2007zk, Marchesano:2009rz, Blumenhagen:2009qh}.
As another possibility, the SM and dark sector can originate from two stacks of (fractional) D3-branes
sitting on two singularities respectively \cite{Aldazabal:2000sa}.
In this case, the number of chiral fermions depends on the structure of the singularities,
and flavor structure depends on the singularity\footnote{
For instance, ${\mathbb C}^3/{\mathbb Z}_3$ singularity can
lead to $SU(3)^3$ gauge group, 3-generation and $SU(3)_{R}$ symmetric Yukawa coupling
$y_{ijk} \propto \epsilon_{ijk}$ at the high energy scale \cite{Cicoli:2012vw}.
Non-commutative deformations of the singularity can also modify the flavor structure.
}
and non-perturbative effects similarly to the previous case \cite{Berg:2012aq}.
Then the MSSM-like models or GUT-like ones may be considered at high energy scales
on respective cycles.
Non-linear realization of supersymmetry (SUSY) or just the SM-like models may be also constructed
with anti-branes or non-supersymmetric fluxes in would-be metastable configurations\footnote{
See discussions on the presence of anti-branes in a supersymmetric flux-background \cite{Kachru:2002gs}.
}.

Secondly, the number of D-branes can be constrained on the CY space even
if one wants to do so freely.
Induced RR charges carried by D-branes wrapping on a cycle, that is,
the tadpoles for RR tensor fields, are required to vanish entirely
for cancelling field-theoretic (mixed) quantum anomalies at low energy scales.
As such induced charges are proportional to the number of the relevant D-branes,
the tadpole conditions can restrict the total number of D-branes wrapping on respective cycles,
i.e. the total rank of the sum of gauge groups in string models.\footnote{
Although in this paper we use the terminology in type II models for simplicity,
one may consider similar situations in heterotic models. The tadpole condition (of the dual six-form potential)
can be given by ${\rm Tr}({\cal R}_2 \wedge {\cal R}_2) - {\rm Tr}(F_2 \wedge F_2) = 0$,
where ${\cal R}_2$ is the geometric curvature two-form on a CY space, $F_2$ is gauge field strength two-form
in $E_8 \times E'_8$ or $SO(32)$,
and we have not included any five-brane contributions.
The total rank of the sum of gauge groups should be less than sixteen,
and thus heterotic models can also contain the hidden sector gauge group \cite{Braun:2013wr}.
}
Focusing on moduli stabilization, orientifolding CY spaces is important.
On such compactifications O-planes carrying negative RR-charges
and hence closed string fluxes can be contained \cite{Giddings:2001yu}.
On top of them, lots of D-branes will be naturally required to cancel the negative
RR-charges, forcing hidden sectors to be contained \cite{Cvetic:2012kj}.
As a consequence,
all string moduli can be stabilized at low energies via combinations of the fluxes
\cite{Douglas:2006es}
and non-perturbative effects on D-branes
\cite{Kachru:2003aw}.
Let us consider, for instance, the D3-brane tadpole condition in F-theory compactified on an elliptically
fibered CY fourfold ($CY_4$)
\beq
N_{\rm D3} + \frac{1}{2}N_{\rm flux} = \frac{\chi (CY_4)}{24}.
\label{tadpoleF}
\eeq
Here $N_{\rm D3}$
denotes the number of D3-branes, $N_{\rm flux}$ is four-form flux contribution to the D3-brane charge,
and $\chi (CY_4)$ is the Euler characteristic of the $CY_4$.\footnote{
In type IIB language, $N_{\rm flux}$ consists of contributions from NSNS-,
RR-fluxes and world volume fluxes on D7-branes,
while the $\chi (CY_4)$ comes from O3-planes, and geometric curvature on O7-planes and D7-branes
\cite{Collinucci:2008pf}.}
In typical cases, r.h.s. of Eq.(\ref{tadpoleF}) is much greater than of ${\cal O}(1)$,
which is the rank of the SM gauge group at low energies.
For instance, one can find sixteen in toroidal orientifolds or of ${\cal O}(10^2 -10^3)$
in more generic cases.
Thus the total number of D3-branes can be naturally split as
$N_{\rm D3} = N_{\rm SM} + N_{\rm para} + \cdots$, where $N_{\rm SM}$ and $N_{\rm para}$
are the rank of the SM gauge group and the dark sector at the high energy respectively,
though Eq.(\ref{tadpoleF}) depends also on $N_{\rm flux}$.
At any rate, thanks to the tadpole condition with O-planes, it would be natural for string theories
to contain multiple SM-like sectors\footnote{
Even if the SM is realized through strong coupling dynamics, say,
Seiberg duality cascade with $N_{\rm SM} \gg 1$ \cite{Seiberg:1994pq},
such a theory would contain also non-trivial hidden sector at low energy scales.
}.

Let us discuss about the reheating of both the SM sector and the parallel world.
An inflaton or closed string moduli which are propagating in the bulk of a CY space
can play a role in reheating both the sectors,
when such states dominate over the energy density in string theory at last.
The states denoted by $\phi$ would be universally coupled to both sectors via gravitational
interaction
\beq
{\cal L} = \phi \left(
{\cal O}_{\rm SM} + {\cal O}_{\rm para}
\right) ,
\eeq
because they exist in the bulk like the graviton. Here we have used in the Planck unit such that
$M_{\rm Pl} = 2.4\times 10^{18}$ GeV $\equiv 1$.
Leptogenesis will work successfully in both sectors when the decay temperature
of $\phi$ is high enough, or when right-handed neutrinos are non-thermally produced
with a sufficient amount in the $\phi$ decays.
Note that if there are other hidden sectors lighter than $\phi$ they will be also produced similarly.
A slight difference between the reheating temperatures in two sectors may come
from different matter contents at high energy scales depending on
brane configurations or topologies in local geometries;
extra states, e.g. heavy gauge bosons or vector-like matters, may be reheated just
in the SM sector or soft SUSY-breaking terms in the dark sector may be much larger
than those in the SM sector.
For instance, a stack of D-branes can sit on an orientifolded local geometry
while an another does not \cite{Blumenhagen:2009gk}.
Alternatively, the SM may localize on a stack of D-branes while the dark sector does on anti-D-branes,
cancelling the tadpole generated by the SM branes.

On the other hand, open string moduli (position) \cite{Tatar:2009jk}
or chiral singlets under the SM gauge group \cite{Blumenhagen:2006xt} may play the role of
right-handed neutrinos.\footnote{
One might consider Kaluza-Klein modes as such neutrinos \cite{Bouchard:2009bu}.
However, for thermal leptogenesis,
we need extremely heavy $\phi$, and then such massive neutrinos are beyond the scope
of four dimensional approximation.
}
For the former states on D7-branes, they will obtain the masses which will be given by
\beq
M_i \sim  f \frac{M_{\rm Pl}}{{\cal V}},~~~
M^\prime_i \sim f^\prime \frac{M_{\rm Pl}}{{\cal V}}
\eeq
via flux compactifications, if exist. Here $f$ and $f'$ are complex flux-parameters, and
${\cal V}$ is the volume of the relevant CY orientifold space in the string unit.
Hence ${\cal V} ={\cal O}(10^7)$ will be required.\footnote{
In such a case, one can get a low reheating temperature via the bulk modulus decay, and hence
a bulk string-theoretic-axion can become a component of dark radiation \cite{Cicoli:2012aq}
as already mentioned production of hidden light modes on branes via $\phi$-decay.
However, strictly speaking, such consequence may depend on moduli stabilization.
}
Depending on the choice of quantized fluxes and flavor structure, one can get $M_1  < M^\prime_1$,
$m_{\nu_3} < m_s$ and a large CP phase.
For the latter states localized between D-branes in similar type IIB orientifolds,
they acquire masses via non-perturbative effects:
\beq
M_i \sim M_{\rm Pl}e^{-{\rm Vol}(E)+ i\vartheta},~~~  M^\prime_i
\sim M_{\rm Pl}e^{-{\rm Vol}(E^\prime) + i\vartheta^\prime} .
\eeq
This is because of respective anomalous $U(1)$ symmetries under which right-handed neutrinos
in respective sectors can be charged.
Here ${\rm Vol}(E)$ and ${\rm Vol}(E^\prime)$ denote the volume of an Euclidean D-branes relevant
to the SM and dark sector, respectively, while $\vartheta$ and $\vartheta^\prime$ are integrands
of RR-fields on the Euclidean branes.
Via the same origin of such an instanton or
a moduli stabilization such that ${\rm Vol}(E) \gtrsim {\rm Vol}(E')$,
$M_1  < M^\prime_1$ and other desired situations may be realized.

\section{Discussion and conclusions}
\label{sec:4}

The BBN bound on the temperature of the parallel world \REF{bbn} implies that the reheating of the
SM and parallel sectors are slightly asymmetric. As already mentioned in the previous section,
the inflaton or closed string moduli propagating in the bulk of a CY space may be responsible
for the asymmetric reheating. Here we briefly mention another possibility. If SUSY is realized
in nature, there will be a fermion that partners with the graviton, called the gravitino, which
has interactions suppressed by either the Planck scale or the SUSY breaking
scale, and therefore is long-lived. Also, the gravitinos are produced in various processes,
such as thermal particle scatterings~\cite{Bolz:2000fu,Pradler:2006qh,Rychkov:2007uq}
and non-thermal production from the decay of moduli~\cite{moduli,Dine:2006ii,Endo:2006tf}
and inflaton~\cite{Kawasaki:2006gs,Asaka:2006bv,Endo:2006qk,Endo:2007ih,Endo:2007sz}.
Thus, it is possible that our Universe was gravitino-rich, or even
gravitino-dominated~\cite{Jeong:2012en}. In the latter case, since the gravitino is coupled to
both the SM and parallel sectors, the two sectors will be heated up to temperatures that are
comparable to each other.
The slight difference of the temperatures could arise if the SUSY particles have mass comparable
to the gravitino and if the masses are slightly different between the two sectors.
We also note that SUSY forbids a sizable coupling between the Higgs doublets
of the SM and parallel sectors, as would be required for temperature asymmetry.

So far we have considered a parallel sector with particle contents and interactions that are identical
to the SM.
In fact, our argument on the relation between $\dnf$ and $N_s$ can be applied to a broader class of
the parallel world. For instance, our result holds even if there is no counterpart of the strong interaction.
It is also straightforward to extend our results to a case with different number of p-neutrino species
and/or p-photons. If it turns out that the relation \REF{pred} is not satisfied, it would imply that
the parallel world is not just a replica of the SM. In any case, if the existence of dark radiation
and/or sterile neutrinos are discovered, it would provide us with a valuable information of the
dark sector.

In this paper we have pointed out that the existence of dark radiation suggested by recent
observations, if confirmed, would pose another coincidence problem of why the density of dark
radiation is comparable to that of photons.
Combined with the long-standing baryon-dark matter coincidence, it may imply that there is a dark
parallel world which is quite similar to the SM. Assuming only gravitational interactions between
the two sectors, we found that there is a rather robust prediction
for the relation between the abundance of dark radiation and the sterile neutrino,
\REF{teq} and \REF{tdiff}.  See Fig.~\ref{fig1} for the relation.
The relation is mostly independent of details of the parallel world, and is simply determined
by the number of light degrees of freedom.
If confirmed, it can serve as a smoking-gun evidence of the parallel world.
We also pointed out that  p-neutrons are a plausible candidate for dark matter, which can have
a cross section of $\sigma_{\rm DM}/m_{\rm DM} = {\cal O}(0.01  - 0.1) \,{\rm cm^2/g}$ 
for the p-QCD scale several times larger than the QCD scale\footnote{
For the pion mass at 700 MeV, 
one obtains 1.6 fm scattering length between two neutrons \cite{Doi:2012ab}.
Then the cross section will fall in the right range.
}.
Interestingly such dark matter is consistent with the observations and may ameliorate the
central density problem~\cite{Rocha:2012jg,Peter:2012jh,Zavala:2012us}. 
We have also discussed the implications
for such parallel world from the string theory point of view. What is important is the presence of small
extra dimensions with rich structures.
The SM and its parallel world can be found through compactifications,
where there are many independent cycles and two sectors can exist on the cycles separately.
Furthermore, consistency conditions to avoid anomalies can also require the parallel world,
being consistent with moduli stabilization.
Bulk modes coupled to the both sectors via gravitation can play a role in reheating these sectors
and hence in generating the origin of baryon asymmetry, e.g., right-handed neutrinos,
whose masses will be generated through flux compactifications or non-perturbative effects in both sectors.
Once the SM sector exists in the string theory, there is no reason why we should not have a similar sector.

\section*{Acknowledgment}
We thank K. Yazaki,  Y. Hidaka and K. Hagino for discussions on nucleon scattering.
This work was supported by the Grant-in-Aid for Scientific Research on Innovative
Areas (No.24111702, No. 21111006, and No.23104008) [FT], Scientific Research (A)
(No. 22244030 and No.21244033) [FT], and JSPS Grant-in-Aid for Young Scientists (B)
(No. 24740135) [FT]. This work was also supported by World Premier International Center Initiative
(WPI Program), MEXT, Japan [FT],
and by Grants-in-Aid for Scientific Research from the Ministry of Education, Science, Sports,
and Culture (MEXT), Japan, No. 23104008 and No. 23540283 [KJS].

\end{document}